\documentclass[a4paper,11pt]{article}
\pdfoutput=1 

\usepackage{jheppub} 

\usepackage[T1]{fontenc} 
\usepackage[utf8]{inputenc} 

\usepackage[dvipsnames]{xcolor}
\usepackage{color,changes}

\title{\boldmath $96$~GeV Scalar Boson in the 2HDM with U$(1)_H$ Gauge Symmetry}

\author[a]{Seungwon Baek,}
\author[b,c]{P. Ko,}
\author[d]{Yuji Omura}

\author[e]{and Chaehyun Yu}

\affiliation[a]{The Institute of Basic Science, Korea University, Anam-ro 145, Seoul 02841, Korea}
\affiliation[b]{School of Physics, Korea Institute for Advanced Study (KIAS), 
85 Hoegi-ro, Seoul 02455, Korea}
\affiliation[c]{Quantum Universe Center (QUC), 
KIAS, 85 Hoegi-ro, Seoul 02455, Korea} 
\affiliation[d]{Department of Physics, Kindai University, Higashi-Osaka, Osaka 577-8502, Japan}
\affiliation[e]{Department of Physics Education and RINS, Gyeongsang National University, Jinju 52828, Korea}

\abstract{
In this paper, we study two Higgs doublet models with gauged $U(1)_H$ symmetry instead of the 
usual softly broken $Z_2$ symmetry, motivated by the excesses around 96 GeV reported by 
the CMS collaboration in the searches for light resonances decaying to two photons and two $\tau$'s.
In this model, one Higgs doublet field is charged under the $U(1)_H$ symmetry,   
thereby one can  avoid  tree-level flavor changing neutral currents. 
The extra $U(1)_H$ gauge symmetry requires extra chiral fermions, in order to satisfy the anomaly-free conditions. 
We analyze the signals of the light resonances, taking into account the contribution of 
the extra fermions, and discuss the consistency with the experimental results in this model.
}


\begin{document} 
\maketitle
\flushbottom


\section{Introduction}
\label{sec:intro}

The Standard Model (SM) has been established as a theory describing particle physics. Almost all predictions of the SM are consistent with experimental results and the Higgs particle was finally discovered at the LHC~\cite{ATLAS:2012yve,CMS:2012qbp}. 
It is certain that there are still large uncertainties in some observables, so that
new physics may exist in the energy region that can be explored at the LHC.
The new model, however, should not drastically modify the SM predictions.
A lot of candidates for new physics have been proposed, motivated by
problems in the SM. The mysteries, especially, concerned with
the origin of the electroweak (EW) scale and the vacuum structure of our universe 
seem to suggest new particles around the EW scale.  Thus, 
some extensions of the SM model that reveal new aspects of our universe  may be confirmed near future. 

Recently, the CMS collaboration has reported some excesses 
in the diphoton channel around 96 GeV mass region, that may suggest 
interesting possibilities of the vacuum structure.
The CMS collaboration has surveyed resonances that decay to two photons,
and reported deviations from the expected signals. 
Based on the data at $\sqrt{s}=$ 8 TeV and 13 TeV with the integrated luminosity of $19.7$ fb$^{-1}$ and $35.9$ fb$^{-1}$, respectively,
the result shows a resonance at $95.3$ GeV with a local significance of $2.8 \sigma$. This can be described by a signal strength,~\cite{CMS:2018cyk}
\begin{equation}
\mu_{\gamma\gamma}^{\rm CMS, previous} = 
\frac{\sigma^{\rm exp}(gg\to \ s \to \gamma\gamma)}
{\sigma^{\rm SM}(gg\to H^{\rm SM-like} \to \gamma\gamma)}
= 0.6\pm 0.2.
\end{equation}
Here $s$ denotes the 96 GeV  scalar boson responsible for the resonance
while the $H^{\rm SM-like}$ is the hypothetical Higgs boson with the mass for the resonance.
The analysis with full data Run 2 data set has been already reported by the CMS collaboration. The result shows an excess with a local significance of $2.9 \sigma$ at $95.4$ GeV. This signal strength for the resonance is~\cite{{CMS:HIG-20-002}} 
\begin{equation}
\mu_{\gamma\gamma}^{\rm CMS} = 
\frac{\sigma^{\rm exp}(gg\to \ s \to \gamma\gamma)}
{\sigma^{\rm SM}(gg\to H^{\rm SM-like}\to \gamma\gamma)}
= 0.33^{+0.19}_{-0.12}.
\end{equation}
The ATLAS collaboration has also reported their analysis for the di-photon channel with the full Run 2 data set and found a mild excess with a local significance of $1.7 \sigma$~\cite{ATLAStalk}
which corresponds to the signal strength of~\cite{Biekotter:2023oen}
\begin{equation}
\mu_{\gamma\gamma}^{\rm ATLAS} = 0.18^{+0.10}_{-0.10}.
\end{equation}
The combined signal strength of the CMS and ATLAS results without possible correlation is~\cite{Biekotter:2023oen}
\begin{equation}
\mu_{\gamma\gamma}^{\rm exp} = 0.24^{+0.09}_{-0.08}
\end{equation}
at the mass of $95.4$ GeV.
The CMS collaboration has also reported another resonance which may be a candidate for an extra scalar boson
in the di-$\tau$ channel. The local significance at 95 GeV is about $2.6 \sigma$ and the corresponding
signal strength is~\cite{CMS:2022goy}
\begin{equation}
\mu_{\tau\tau}^{\rm CMS} = 
\frac{\sigma^{\rm exp}(gg\to \ s \to \tau\tau)}
{\sigma^{\rm SM}(gg\to H^{\rm SM-like}\to \tau\tau)}
= 1.2\pm 0.5,
\end{equation}
while there has been no analysis at ATLAS for the corresponding region so far.
In addition, the LEP collaboration has announced a resonance at a similar invariant mass of a $b\bar{b}$ pair.
The resonance in the $e^+ e^- \to Z s \to Z b\bar{b}$ channel was observed with a local significance of $2.3 \sigma$.
The excess can be interpreted as the signal strength of~\cite{Cao:2016uwt,Azatov:2012bz}
\begin{equation}
\mu_{b\bar{b}}^{\rm LEP} = 
\frac{\sigma^{\rm exp}(e^+ e^-\to Z (s \to b\bar{b}))}
{\sigma^{\rm SM}(e^+ e^-\to Z (H^{\rm SM-like}\to b\bar{b}))}
= 0.117\pm 0.057,
\end{equation}
where the mass of the resonance is about 98 GeV.

If those deviations are originated from new resonances that reside around
96 GeV, one of the good candidates is a neutral scalar originated from scalar fields that contribute 
to the EW symmetry breaking 
\cite{Biekotter:2023oen,Biekotter:2019kde,Biekotter:2019mib,Biekotter:2020cjs,Biekotter:2021qbc,Heinemeyer:2021msz,
Biekotter:2022jyr,Biekotter:2023jld,Bhattacharya:2023lmu,Banik:2023ntf}. 
The possibility that extra scalar fields may exist has been discussed in many works. 
Adding extra scalar fields is surely one simple way to extend the SM without disturbing the anomaly-free conditions. Scalar fields may play a role in breaking extra gauge symmetries.
In fact, extra neutral scalars that are related to gauge symmetry breaking and
reside around 96 GeV have been proposed motivated by the excesses, and the signals have been studied
in many works \cite{Kundu:2019nqo,Cao:2019ofo,Abdelalim:2020xfk,Belyaev:2023xnv,Aguilar-Saavedra:2023tql,Dutta:2023cig,Ellwanger:2023zjc,Cao:2023gkc,Arcadi:2023smv,Mulaudzi:2023khg,Liu:2024cbr,Cao:2024axg,Kalinowski:2024uxe,Ellwanger:2024txc,Ellwanger:2024vvs,Arhrib:2024wjj,Benbrik:2024ptw,Lian:2024smg,Khanna:2024bah,BrahimAit-Ouazghour:2024img,Ashanujjaman:2023etj,Banik:2023vxa,Coloretti:2023yyq,Dong:2024ipo,Borah:2023hqw,Ahriche:2023hho,Ahriche:2023wkj,Chen:2023bqr,Dev:2023kzu,Wang:2024bkg,YaserAyazi:2024hpj,Gao:2024qag,Haisch:2017gql,Fox:2017uwr,Biekotter:2021ovi,Azevedo:2023zkg}.

One of the simple extensions of the SM with extra scalar fields   
contributing to the EW symmetry breaking is  two Higgs doublet models (2HDMs). 
In generic 2HDMs, both of the Higgs doublets  
can contribute simultaneously to fermion masses 
when they develop nonzero vacuum expectation values (VEVs). 
Since the fermion mass matrices and Yukawa couplings are not simultaneously diagonalizable, 
Higgs mediated flavor-changing neutral currents (FCNCs) will be present even at tree levels. 
Then a large portion of the parameter space in the generic 2HDMs would have already been excluded by stringent 
constraints from flavor physics such as $M^0 - \overline{M^0}$ mixings (with $M=K, B_d, B_s$), $B \to X_s \gamma$,
$B_s \rightarrow \mu^+ \mu^-$, $l \rightarrow l' \gamma, l \rightarrow  3 l^{'}$ etc., to name a few.

The Higgs-mediated FCNC problem in generic 2HDMs can be cured 
{\it a la} by Natural Flavor Conservation Criterion by Glashow and Weinberg \cite{Glashow:1976nt}.
A simple realization of their criterion is to introduce $Z_2$ symmetry under which two Higgs doublets has different $Z_2$ parity.
Then the SM chiral fermions will also carry different $Z_2$ charges.
There are four different charge assignments for which Yukawa couplings of the SM fermions are allowed 
in such a way that the fermions of the same electric charge get their masses
from one and the same Higgs doublet, either $H_1$ or $H_2$, but not from both. 
Namely 2HDMs that do not predict tree level FCNCs
are classified into four types.
In the Type-I 2HDM, only one Higgs field is coupled to
the SM fermions. In the Type-II 2HDM, one Higgs field ($H_1$) couples with 
right-handed down-type quarks and charged leptons, and the other Higgs field ($H_2$)
couples with right-handed up-type quarks and neutrinos.
It is possible to consider a model where the couplings are flipped:
$H_1$ couples with right-handed down-type quarks and neutrinos, and the $H_2$ couples with right-handed up-type quarks and charged leptons. 
We call this model the Type-Y 2HDM. 
In the fourth 2HDM, $H_1$ couples with quarks and $H_2$ couples with leptons. 
Then this $Z_2$ symmetry is assumed to be softly broken by dim-2 operator, $m_{12}^2 H_1^\dagger H_2 + H.c.$, in order to increase the masses of the scalars originated from $H_1$ and $H_2$. 

In this paper, we consider gauged $U(1)_H$ symmetry instead of softly broken $Z_2$ symmetry, 
by assigning a non-vanishing $U(1)_H$ charge to $H_2$ \cite{Ko:2012hd,Ko:2013zsa,Ko:2014uka,Ko:2015fxa}.
In addition, extra complex singlet scalar field, $\Phi$, is also introduced in order
to break $U(1)_H$ spontaneously. As discussed in Ref. \cite{Ko:2012hd}, there are several possible setups in 2HDMs with gauged $U(1)_H$. 
In this work, we concentrate on the Type-II 2HDM and the Type-Y 2HDM motivated by the excesses around 96 GeV.
In Sect.~\ref{sec2}, we discuss the setups of our 2HDMs.
In Sect.~\ref{sec3}, we summarize experimental constraints relevant to
our models. In Sect.~\ref{sec4}, the analyses concerned with the excesses are shown. Section~\ref{sec5} is devoted to summary. In Appendix \ref{bounded}, the constraints from the condition for the scalar potential to be bounded from below is introduced.
In Appendix \ref{effcoup}, the relevant effective couplings are summarized.

\section{The 2HDM with gauged $U(1)_H$ symmetry}
\label{sec2}

We consider 2HDMs with extra $U(1)_H$ gauge symmetry, 
where one of the Higgs fields, $H_2$, is charged under the $U(1)_H$ gauge symmetry, 
while the other $H_1$ is not charged.
Some SM fermions may be also charged under $U(1)_H$ 
so that phenomenologically viable Yukawa couplings for the SM fermions 
are allowed by the presumed gauge symmetry. 
Below, we discuss the detail based on Refs. \cite{Ko:2012hd,Ko:2013zsa,Ko:2014uka,Ko:2015fxa}.

Due to the charge assignment to the scalar doublet fields, both $H_1^\dagger H_2$ and $(H_1^\dagger H_2)^2$ terms are forbidden in the model. 
It has been found that this setup for the scalar potential is not excluded by 
experiments so far \cite{Jung:2023ckm}.
However, because there is no new scale in the potential, the masses of the new scalar bosons in the model,
in particular, the mass of the charged Higgs boson, are of at most EW scale
of the VEV $v\sim 246$ GeV.
We add a new singlet scalar field $\Phi$ to the model,  which gives rise to the $H_1^\dagger H_2$ term
after symmetry breaking. Then both the charged Higgs and extra neutral Higgs bosons can be heavier than the SM Higgs boson. 

\begin{table}[h]\centering
	\begin{tabular}{|c||c|c|c|}\hline
		& $H_1$ & $H_2$ & $\Phi$ \\ \hline \hline 
		$SU(3)_c$ &{\bf 1} & {\bf 1}& {\bf 1} \\ \hline	
		$SU(2)$  & {\bf 2} & {\bf 2} & {\bf 1} \\ \hline	
		$U(1)_Y$ & $1/2$ & $1/2$ & $0$   \\ \hline	
		$U(1)_H$  & $0$ & $1$ & $-1$ \\ \hline	
	\end{tabular}
	\caption{\label{scalarcharge} The charge assignment to the scalar fields. Irreducible representations of $SU(3)_c$ and $SU(2)$ are denoted by their dimensions in boldface.
	}
\end{table}

The charge assignments under $G_{SM} \times U(1)_H$ gauge symmetry to the scalar fields are shown in Table~\ref{scalarcharge}.
Then the renormalizable parts of the scalar potential in our model is 
\begin{eqnarray}\label{Eq.Pot}
		V(H_k, \Phi) & = & 
		m_k^2 H_k^\dagger H_k + {\lambda_k} ( H_k^\dagger H_k)^2 
+ \lambda_3 ( H_1^\dagger H_1 )( H_2^\dagger H_2 ) + \lambda_4 | H_1^\dagger H_2 |^2  
\nonumber \\
		& & + \ m_\Phi^2 | \Phi |^2 + \lambda_\Phi | \Phi |^4 + \tilde{\lambda}_k | \Phi |^2
		H_k^\dagger H_k - \left\{ {\sqrt{2} \mu_\Phi H_1^\dagger H_2 \Phi }+ H.c. \right\}, 
\end{eqnarray}
where $k=1,2$. 
We emphasize that, compared to the usual 2HDMs, two operators are missing in 
our model due to the $U(1)_H$ gauge symmetry: the soft $Z_2$ breaking dim-2 operator, $m_{12}^2 H_1^\dagger H_2$,   
and the $\lambda_5  (H_1^\dagger H_2)^2 + h.c.$ term. 
Still the $H_1^\dagger H_2 + h.c.$ term can be 
realized from the $H_1^\dagger H_2 \Phi+ h.c$ terms after the singlet field 
$\Phi$ develops a non-vanishing VEV.

The scalar fields are expanded around nonzero VEVs with only $SU(3)_C \times U(1)_{\rm em}$ 
remaining unbroken:
\begin{equation}\label{VEV}
\langle H_k \rangle = ( 0 , v_k /\sqrt{2} )^\intercal , \ \ \langle \Phi \rangle = v_\Phi / \sqrt{2} ,
\end{equation}
and
\begin{equation}
H_k = \left( \begin{array}{c}
\phi_k^+ \\ 
\frac{v_k}{\sqrt{2}} + \frac{1}{\sqrt{2}} ( h_k + i \chi_k^0 ) 
\end{array}
\right) , \quad 
\Phi = \frac{1}{\sqrt{2}} (  v_\Phi + h_\Phi + i \chi_\Phi ),
\end{equation}
where $v_1= v \cos\beta$, $v_2 = v \sin\beta$, and $v=\sqrt{v_1^2+v_2^2}=246$ GeV. 

After the electroweak (EW) and $U(1)_H$ gauge symmetry breaking, 
three CP-even neutral scalar fields, $ h_i \equiv (h_1, h_2, h_\Phi)$, mix among themselves,
and the $3\times 3$ mass matrix is
diagonalized by three physical states, $S_i \equiv (\tilde{h}, h, H)$.
$h$ is identified as the Higgs boson with the mass of 125 GeV while $\tilde{h}$ is as the $96$ GeV scalar boson.
The remaining $H$ is an additional scalar boson. 
The mixing between three scalar bosons is defined by 
\begin{equation}
{S_i = R_{ij} h_j},
\end{equation}
where the rotation matrix $R_{ij}$ is given by
\begin{equation}
R = \left( 
\begin{array}{ccc}
1 & 0 & 0 \\
0 & \cos \alpha_3 & \sin \alpha_3 \\
0 & -\sin \alpha_3 & \cos \alpha_3
\end{array}
\right)
\left( 
\begin{array}{ccc}
\cos \alpha_2 & 0  & \sin \alpha_2 \\
0 & 1 & 0 \\
 -\sin \alpha_2 & 0 & \cos \alpha_2
\end{array}
\right)
\left( \begin{array}{ccc}
\cos \alpha_1 & \sin \alpha_1 & 0 \\
-\sin \alpha_1 & \cos \alpha_1 & 0 \\
0 & 0 & 1 \\
\end{array}
\right)
\end{equation}
with three mixing angles, $\alpha_i$ $(i=1,2,3)$.

The CP-odd states, $\chi = (\chi_\Phi, \chi_1, \chi_2)$,  also mix among themselves and
yield two Nambu-Goldstone bosons, $G_1^0$ and $G_2^0$, and one pseudoscalar boson, $A$.
The mixing angles are defined by $G_i = (V_A)_{ij} \chi_j$, where $ G = (G_1^0, G_2^0, A)$ and
\begin{equation}
V_A = 
\left( 
\begin{array}{ccc}
0 & \cos\beta & \sin\beta \\
\cos\delta & \sin\beta\sin\delta & -\cos\beta \sin\delta \\
\sin\delta & -\sin\beta \cos\delta & \cos\beta \cos\delta
\end{array}
\right)
\end{equation}
with 
\begin{equation}
\cos\delta = \frac{v_\Phi}{\sqrt{v_\Phi^2 + (v\cos\beta \sin\beta)^2}}.
\end{equation}
Then the mass of the pseudoscalar boson is obtained by the mixing matrix as
\begin{equation}
m_A^2 = \mu_\Phi 
\left( 
\frac{v_1 v_\Phi}{v_2} 
+\frac{v_1 v_2}{v_\Phi}
+\frac{v_2 v_\Phi}{v_1}
\right).
\end{equation}
The three neutral gauge bosons mix among them after EW and $U(1)_H$ symmetry breaking, and
their mass matrix is given by 
\begin{equation}
M^2 = \frac{v^2}{8}
\left( 
\begin{array}{ccc}
g^2 & -g g_Y^2 & -2 g g_X s_\beta^2 \\
-g g_Y & g_Y^2 & 2 g_Y g_X s_\beta^2 \\
-2 g g_X s_\beta^2 & 2 g_Y g_X s_\beta^2 & 4 g_X^2 \left(s_\beta^2 + \frac{v_\Phi^2}{v^2} \right)
\end{array}
\right),
\end{equation}
where $g, g_Y$, and $g_X$ are gauge couplings for $U(1)_Y$, $SU(2)_L$, and $U(1)_H$, respectively,
and $s_\beta = \sin\beta$. A massless state is identified as the photon and the remaining two states mix with each other.
Then the two massive bosons, $Z$ and $Z^{'}$, are expressed as
\begin{eqnarray}
 \hat{Z} &=& Z \cos \theta +  Z' \sin \theta, 
\\
\hat{Z}' &=&  Z \sin \theta -  Z' \cos \theta,
\end{eqnarray}
where the $\hat{Z}$ and $\hat{Z}'$ are two states after identifying the photon state.
The mixing angle is defined by
\begin{equation}
\sin \theta = \frac{\lambda}{(1+\lambda^2)^{1/2}},
\end{equation}
where
\begin{eqnarray}
\lambda &=&
\frac{\bar{g}^2 - g_H^2 + \left[ (\bar{g}^2 - g_H^2)^2 + 4 \bar{g}^2 \bar{\delta}^2 \right]^{1/2}}
{2\bar{g}\delta}, \\
\bar{g}^2 &=& g_Y^2 + g^2, \\
g_H^2 &=& 4 g_X^2 \left( s_\beta^2 + \frac{v_\Phi^2}{v^2} \right), \\
\bar{\delta} &=& 2 g_X s_\beta^2.
\end{eqnarray}

\begin{table}[t]\centering
	\begin{tabular}{|c||c|c|c|c|c|c|}\hline
		& $Q^i_L$ &$u^i_{R}$ &$d^i_R$ &$L_L^i$  &	$e^i_R$ &$\nu^i_R$  \\ \hline \hline 
		$SU(3)_c$ & {\bf 3}& {\bf 3}& {\bf 3}& {\bf 1}& {\bf 1}& {\bf 1}  \\ \hline	
		$SU(2)$ & {\bf 2}& {\bf 1}& {\bf 1}& {\bf 2}& {\bf 1}& {\bf 1}  \\ \hline	
		$U(1)_Y$ & $1/6$ &$2/3$ &$-1/3$ &$-1/2$ &$-1$ &$0$    \\ \hline	
		$U(1)_H$ & $-1/3$& $2/3$& $-1/3$& $0$& $0$& $1$  \\ \hline	
	\end{tabular}
	\caption{\label{charge2} The charge assignment to the SM fermions in the Type-II 2HDM.
	}
\end{table}

\begin{table}[t]\centering
	\begin{tabular}{|c||c|c|c|c|c|c|}\hline
		& $Q^i_L$ &$u^i_{R}$ &$d^i_R$ &$L_L^i$  &	$e^i_R$ &$\nu^i_R$  \\ \hline  \hline
		$SU(3)_c$ & {\bf 3}& {\bf 3}& {\bf 3}& {\bf 1}& {\bf 1}& {\bf 1}  \\ \hline	
		$SU(2)$ & {\bf 2}& {\bf 1}& {\bf 1}& {\bf 2}& {\bf 1}& {\bf 1}  \\ \hline	
		$U(1)_Y$ & $1/6$ &$2/3$ &$-1/3$ &$-1/2$ &$-1$ &$0$    \\ \hline	
		$U(1)_H$ & $0$& $1$& $0$& $0$& $-1$& $0$  \\ \hline	
	\end{tabular}
	\caption{\label{chargeY} The charge assignment to the SM fermions in the Type-Y 2HDM.
	}
\end{table}

The fermion sector depends on the $U(1)_H$ charge assignments to the SM fermions. 
It turns out that the Type-I 2HDM for the fermions can be constructed by adding right-handed neutrinos 
without any gauge anomalies \cite{Ko:2012hd,Ko:2013zsa,Ko:2014uka}, while
models of other types require extra fermions as well as right-handed neutrinos in order to cancel gauge 
anomalies \cite{Ko:2012hd,Ko:2015fxa}.
In this paper we focus on the Type-II and Type-Y models which are relevant to the 96 GeV scalar resonance 
signals \cite{Biekotter:2019kde}. 
In Tables \ref{charge2} and \ref{chargeY}, we show the $U(1)_H$ charge assignments to the SM fermions 
in the Type-II and Type-Y 2HDMs, respectively.

Then we obtain the same Yukawa interactions as those in the usual 2HDM in both types.
In the Type-II 2HDM, the Yukawa interaction is 
\begin{equation}
  {\cal L}_{\rm Yukawa}^{II}=-  Y^{ij}_u \overline{Q^i_L} \widetilde H_2 u_R^j- Y^{ij}_d  \overline{Q^i_L} H_1 d_R^j  
  - Y^{ij}_e \overline{L^i_L} H_1  e_R^j - Y^{ij}_n \overline{L^i_L} \widetilde  H_2  \nu_R^j+h.c.,
\end{equation}
while in the Type-Y 2HDM the Yukawa interaction is given by
\begin{equation}
  {\cal L}_{\rm Yukawa}^{Y}=-  Y^{ij}_u \overline{Q^i_L} \widetilde H_2 u_R^j- Y^{ij}_d  \overline{Q^i_L} H_1 d_R^j  
  - Y^{ij}_e \overline{L^i_L} H_2  e_R^j - Y^{ij}_n \overline{L^i_L} \widetilde  H_1  \nu_R^j+h.c.,
\end{equation}
where, $\widetilde H_{1,2} = i \sigma_2 H_{1,2}^*$ and $i , j=1,\,2,\,3$.

In both models, the gauge anomalies involving $U(1)_H$ currents are not cancelled with SM fermions only, 
and additional new chiral fermions have to be introduced in order to fulfill the anomaly cancellation conditions. 
We could find a lot of setups without gauge anomaly \cite{Ko:2012hd}. 
The Type-II 2HDM, for instance, can be realized by the model inspired by the $E_6$ Grand Unified Theory \cite{Ko:2015fxa}. The anomaly-free conditions in the Type-Y 2HDM can be satisfied,
by adding extra quarks and leptons in Table \ref{chargeY2} when the $U(1)_H$ charge assignments to the 
SM fermions are as in Table \ref{chargeY}. 
Note that Yukawa couplings between $\Phi$ and extra fermions are allowed by the 
assumed gauge symmetries:
\begin{equation}
 {\cal L}^Y_{{\rm extra}}= -y^{i}_D \overline{u^{\prime \, i}_L} \Phi^\dagger u^{\prime \, i}_R -y^{i}_L \overline{e^{\prime \, i}_L} \Phi e^{\prime \, i}+h.c..
\end{equation}
\begin{table}[t]\centering
	\begin{tabular}{|c||c|c|c|c|}\hline
		& $u^{\prime \, i}_{ L}$ &$u^{\prime \, i}_{ R}$ &$e^{\prime \, i }_L$ &$e^{\prime \, i }_R$   \\ \hline  \hline
		$SU(3)_c$ & {\bf 3}& {\bf 3}& {\bf 1}& {\bf 1} \\ \hline	
		$SU(2)$ & {\bf 1}& {\bf 1}& {\bf 1}& {\bf 1}  \\ \hline	
		$U(1)_Y$ & $2/3$ &$2/3$ &$-1$ &$-1$    \\ \hline	
		$U(1)_H$ & $1$& $0$& $-1$& $0$ \\ \hline	
	\end{tabular}
	\caption{\label{chargeY2} The charge assignment to the extra fermions in the Type-Y model.
	}
\end{table}
In this paper, we do not pay attention to the extra fermions too much because they strongly depend on 
the model construction which is not unique. Our motivation is to study the effects of the extra fermions
to the scalar boson decays such as $\tilde{h} \rightarrow \gamma\gamma$ through the loops. 
Thus, the extra fermion contributions to the loop diagrams must be considered collectively, that is the sum 
of all the fermions in the model, which is quite cumbersome.
Instead, we assume a vectorlike charged lepton and quark as a representative of collection of the extra fermions in the loop, denoting the Yukawa couplings of extra quarks and leptons as $y_D$ and $y_L$, respectively. 
In our phenomenological analysis, those parameters are fixed at $y_D=y_L=1$.

\section{Constraints}
\label{sec3}

In this section, we consider various theoretical and phenomenological 
constraints on our models.

\subsection{Theoretical constraints}

First, we consider the constraints from the vacuum stability conditions for 
nonzero VEVs.
The requirement for the scalar potential to be stable and bounded from below 
constrains the dimensionless couplings in the scalar potential.
Following the approach in Ref. \cite{Arhrib:2011uy}, we find the constraints on the dimensionless couplings,
which are presented in appendix \ref{bounded}.

The models also have constraints from perturbative unitarity bounds. Following the method to find the constraints from perturbative unitarity \cite{Horejsi:2005da,Muhlleitner:2016mzt},
we find that the constraints are given by
\begin{eqnarray}
|b_\pm|, |c_\pm|, |f_\pm|, |g_\pm|, |f_{s,s_1,s_2}|, \frac{1}{2}|a_{1,2,3}| \le 8\pi,
\end{eqnarray}
where
\begin{eqnarray}
b_\pm &=& 2\lambda_{1,2}, \quad c_{\pm}=\lambda_1+\lambda_2\pm\sqrt{(\lambda_1-\lambda_2)^2
+\lambda_4^2},
\nonumber \\
f_\pm &=& \lambda_3 +\lambda_4\pm \lambda_4, \quad
g_\pm = \lambda \pm \lambda_4,
\nonumber \\
f_s &=& 2 \lambda_\Phi, \quad f_{s_1} = \tilde{\lambda}_1, \quad 
f_{s_2} = \tilde{\lambda}_2,
\end{eqnarray}
and $a_{1,2,3}$ are roots of the cubic equation for $x$:
\begin{eqnarray}
0&=&x^3  -4(3\lambda_1+3\lambda_2+2\lambda_\Phi)x^2
\nonumber \\
&-&4(2\tilde{\lambda}_1^2+2\tilde{\lambda}_2^2  + 36 \lambda_1 \lambda_2
-4\lambda_3^2 - 4 \lambda_3\lambda_4 -\lambda_4^2
+24\lambda_1 \lambda_\Phi + 24 \lambda_2 \lambda_\Phi) x
\nonumber \\
&+&
96 \tilde{\lambda}_1^2\lambda_2 + 96 \tilde{\lambda}_2^2 \lambda_1
-64 \tilde{\lambda}_1\tilde{\lambda}_2\lambda_3
-1152\lambda_1\lambda_2\lambda\Phi
+128\lambda_3^2 \lambda_\Phi + 128 \lambda_3\lambda_4\lambda_\Phi
+32\lambda_4^2\lambda_\Phi.
\nonumber
\end{eqnarray}

\subsection{Electroweak precision observables}

In order to study the physical effects  on the electroweak precision observables 
(EWPOs) in this model, full calculations of the relevant amplitudes at the one-loop level are required, 
which are quite involved.  Instead, we calculate the $\Delta T$ parameter approximately.
Due to the $Z$-$Z'$ mixing, the mass of the $Z$ boson is shifted 
\begin{equation}
m_Z^2 = \frac{m_W^2}{c_W^2}{\cos^2 \theta} - m_{Z'}^2\frac{\sin^2 {\theta}}{\cos^2 {\theta}},
\end{equation}
at tree level. We note that the $W^\pm$ mass is not shifted by the $U(1)_H$ gauge symmetry.
Since the mixing angle is expected to be small, this mass shift changes the $\rho$ parameter
\begin{equation}
\Delta \rho_{\theta} = 1-\frac{1}{\rho} \approx
-\sin^2 \theta \left(1-\frac{c_W^2 m_{Z'}^2}{m_W^2}\right)
\end{equation}
up to the leading order of $s_\theta$. This is converted to the $T$ parameter as
$T_{\theta} = \Delta \rho_{\theta} /\alpha(m_Z)$.

The new scalar bosons also contribute to the shift of the $\rho$ parameter at loop levels.
At one-loop level, we find that the approximate formula of the contribution is \cite{Grimus:2007if}
\begin{eqnarray}
T_s &=&
\frac{1}{16 \pi m_W^2 s_W^2}
\left[ 
g_{W^{\pm}H^{\mp} A}^2 F(m_{H^\pm}^2, m_A^2)
+\sum_i g_{W^{\pm}H^{\mp} S_i}^2 F(m_{H^\pm}^2, m_{S_i}^2)
\right.
\nonumber \\
&&
\qquad\qquad\quad
-\sum_i g_{Z A S_i}^2 F(m_{A}^2, m_{S_i}^2)
+3\sum_i g_{Z Z S_i}^2 F(m_{Z}^2, m_{S_i}^2)
\nonumber \\
&&
\qquad\qquad\quad
-3\sum_i g_{WW S_i}^2 F(m_{W}^2, m_{S_i}^2)
+3\sum_i g_{Z Z' S_i}^2 F(m_{Z'}^2, m_{S_i}^2)
\nonumber \\
&&
\qquad\qquad\quad
-3 g_{W^\pm H^\mp Z'}^2 F(m_{Z'}^2, m_{H^\pm}^2)
-3 g_{W^\pm H^\mp Z}^2 F(m_{Z}^2, m_{H^\pm}^2)
\nonumber \\
&&
\left.
\qquad\qquad\quad
-3  F(m_{Z}^2, m_{h}^2)
+3 F(m_{W}^2, m_{h}^2)
\right],
\end{eqnarray}
where the loop function $F(x,y)$ is defined by
\begin{equation}
F(x,y) = \frac{x+y}{2}-\frac{xy}{x-y} \ln \frac{x}{y}
\end{equation}
for $x\neq y$ while $F(x,y)=0$ for $x=y$.
The effective couplings, $g_{W^{\pm}H^{\mp} A}$ and so on, are defined in Appendix~\ref{effcoup}.
This loop contribution to $T$ in the case of 2HDMs with $Z_2$ symmetry vanishes.
However, in our model, the EW symmetry breaking makes the $Z'$ boson massive by
absorbing one of degrees of freedom in the scalar fields. Therefore, the presence of 
the $U(1)_H$ gauge symmetry implies that the contribution from the $Z$-$Z'$ mixing must be taken into consideration
together with the scalar loop corrections.
Finally, the sum of two contributions leads to
\begin{equation}
T = T_\theta + T_s,
\end{equation}
where the measured value is 
\cite{ParticleDataGroup:2024cfk}
\begin{equation}
T^{\rm exp} = 0.04 \pm 0.03.
\end{equation}
We constrain the $T$ parameter within $2\sigma$.

\subsection{Flavor physics}

The charged Higgs boson mass $m_{H^\pm}$ and $\tan\beta$ are strongly constrained by flavor physics, 
in particular, by the $b \to s\gamma$ decay.
The constraints on the 2HDMs from flavor physics including the experimental results at the LHC   
were analyzed in Ref. \cite{Arbey:2017gmh,Haller:2018nnx}.
In the Type-II 2HDM, the lower bound of the charged Higgs mass is about 600 GeV, and this is almost independent of $\tan \beta$. The study of $B \to X_s \gamma$ in Ref. \cite{Misiak:2017bgg} also has led the same lower bound. 
In Ref. \cite{Misiak:2020vlo}, the authors improve the SM prediction of $B \to X_s \gamma$: $Br(B \to X_s\gamma) = (3.40 \pm 0.17) \times 10^{-4}$ \cite{Misiak:2020vlo}.
They also comment that the bound of $m_{H^\pm}$ becomes very severe:
$m_{H^\pm} \gtrsim 800$ GeV \cite{Misiak:2020vlo}.\footnote{See also Ref. \cite{Li:2024kpd}. }
One of the reasons why the bound becomes so strong is that
the improved SM prediction is much higher than the experimental result, $Br(B \to X_s\gamma) = (3.32 \pm 0.15) \times 10^{-4}$ \cite{HFLAV:2016hnz}.
In the Type-II 2HDM, the charged Higgs contribution to $B \to X_s \gamma$
interferes with the SM contribution in a constructive manner,
so that such new physics contribution is strongly limited. 

Recently, the  Heavy Flavor Averaging Group (HFLAV) announced the new result, $Br(B \to X_s\gamma) = (3.49 \pm 0.19) \times 10^{-4}$ \cite{HeavyFlavorAveragingGroupHFLAV:2024ctg}, that is a little larger than the SM prediction. 
If the latest HFLAV result is used, the $m_{H^\pm}$ bound is expected to be relaxed.
We calculate $Br(B \to X_s\gamma)$ using SuperIso v4.1 \cite{Mahmoudi:2008tp}
and compare the new HFLAV result. 
We derive the lower bound of the charged Higgs mass as $m_{H^\pm} \gtrsim 500$ GeV, requiring that our prediction is within $2 \sigma$ of the new HFLAV result. The more detailed study is beyond our scope, but 
more accurate analysis of $B \to X_s\gamma$ may improve the bound.

Since both bounds are meaningful, we present plots for two cases: $m_{H^\pm} > 500$ GeV and $m_{H^\pm} > 800$ GeV.
We refer to the former bound as the HFLAV cut and the latter as the tight cut.
We note that the bound from $R_\gamma = Br_{(s+d)\gamma}/Br_{cl\bar{\nu}}$ is stronger than that from
$Br(B\to X_s\gamma)$~\cite{Li:2024kpd}.
Therefore, for the tight cut, we apply the constraints on $m_{H^\pm}$ and $\tan\beta$ as provided in Ref.~\cite{Li:2024kpd},
where both bounds are simultaneously considered.
However, the HFLAV group has not yet reported a value for $R_\gamma$. 
In Ref.~\cite{Li:2024kpd}, the bounds from $Br(B\to X_s \gamma)$ and $R_\gamma$ differ by about 50 GeV 
in the lower limit on $m_{H^\pm}$.
Taking this difference into account, we present plots using the constraint $m_{H^\pm} > 500$ GeV for the HFLAV cut.
In addition, we have tested our models with a more conservative constraint, $m_{H^\pm} > 550$ GeV, 
motivated by the bound from $R_\gamma$.
However, we find that the resulting behavior is not significantly affected by the 50 GeV difference in the $m_{H^\pm}$ bound.

Another stringent constraint from flavor physics arises from the $B_s\to \mu^+ \mu^-$ decay.
In the Type-II 2HDM, this constraint is significant
in both the low and high $\tan\beta$ regimes, {\it i.e.}, $\tan\beta \sim 1$ and $\tan\beta \sim 20$. 
In contrast, in the Type-Y 2HDM, only the low $\tan\beta$ regime is affected. 
We apply these constraints to both the HFLAV and tight cuts as provided in Ref.~\cite{Li:2024kpd}.
In our model,
there are extra quarks that may lead destructive interference with the SM one.
The study of flavor physics concerned with the extra quarks would not be simple because 
the other flavor observables as well as the collider constraints come into play. 
The detailed study of flavor physics is our future work.

\subsection{Experiments for scalar bosons}

There have been a lot of searches for new scalar bosons
at the LEP, Tevatron, and LHC, and lots of data have been accumulated. 
However, there is no definite signal for a new scalar boson so far.
These search data except for the resonance data around $96$ GeV will strongly constrain our model.
We apply these experimental constraints for the additional scalar bosons using the public code {\tt HiggsBounds} \cite{Bechtle:2020pkv} which is encoded in {\tt HiggsTools}~\cite{Bahl:2022igd}.

In our model, the 125 GeV scalar boson will mix with other neutral scalar bosons.
Therefore, these mixings change the predictions for the properties of the observed Higgs boson.
Then the precision observables for productions and decays of the 125 GeV Higgs boson also 
constrain our model strongly. 
We use the public code {\tt HiggsSignals} \cite{Bechtle:2020uwn} to test the 125 GeV 
Higgs boson in our model. In analysis, we utilize {\tt HiggsTools}~\cite{Bahl:2022igd} which contains the most recent version of {\tt HiggsSignals}.
In order to ensure that the 125 GeV Higgs boson in our model mimics the behavior of the SM Higgs boson, we impose the conditions
$\Delta \chi_{125}^2 < 6.18$ and $\Delta \chi^2 < 0$~\cite{Chen:2023bqr},
where $\Delta \chi_{125}^2$  is the difference of $\chi^2$ for the measurement of the 125 GeV Higgs boson between our model and the SM,
while $\Delta \chi^2 $ is that of the combined $\chi^2$ of the measurements at the 96 GeV excess as well as those
of the 125 GeV Higgs Boson.

\subsection{Constraints from $t\bar{t}\tau^+\tau^-$ production at the LHC}
\label{sec4-3}
The 96 GeV scalar boson couples to both the top quark and tau lepton. Thus, the production of $t\bar{t}\tau^+\tau^-$ at the LHC
can constrain the scalar boson \cite{Iguro:2022dok}\footnote{See also Ref. \cite{Iguro:2022fel}.}. 
Based on the ATLAS Run 2 full data \cite{ATLAS:2022yrq}, the authors in Ref. \cite{Iguro:2022dok}
have proposed the upper bounds of 
the Yukawa couplings, $\rho^{tt, \tau\tau}_{\tilde{h}}$, normalized by  $\sqrt{2}$.
Recently, the CMS collaboration also announced the result of the new physics search in the $t\bar{t}\tau^+\tau^-$ channel \cite{CMS:2024ulc}, and the upper bound of the cross section is about $0.6$ times lower than those in Ref.\cite{Iguro:2022dok}. Using the CMS results, we require that the Yukawa couplings
should satisfy 
\begin{eqnarray}
1.07 \times \left | \rho_{\tilde{h}}^{tt} \right |^2 { \rm Br} \left ( \widetilde h \to \tau \overline{\tau} \right ) &<0.03& ,
\label{tttautau}%
\end{eqnarray}
where $\rho_{\tilde{h}}^{tt} =\frac{g_{\tilde{h}}^{t, \textrm{Y}} m_t}{v} $ is defined.
It is worthwhile to mention that these constraints are based on the analysis
of the tree-level calculation for the $t\bar{t}\tau^+\tau^-$ production \cite{Iguro:2022dok}. 
With QCD corrections to the process, the bounds may be slightly modified.

\section{Results}
\label{sec4}

In this section, we analyze the resonance signals in both the Type-II 2HDM and the Type-Y 2HDM.

\subsection{Parameters}
\label{sec4-1}

In our model, there are 11 parameters as follows:
\begin{equation}
\alpha_1, \alpha_2, \alpha_3, \tan\beta, v_\Phi, m_H, m_A, m_{H^\pm}, y_D, y_L, g_X,
\end{equation}
which denote three mixing angles between CP-even neutral Higgs bosons, $v_2/v_1$, the VEV of $\Phi$,
the heavy neutral Higgs boson mass, the pseudoscalar boson mass, the charged Higgs boson mass,
the Yukawa couplings of extra quark and lepton, and the gauge coupling of the $U(1)_H$ symmetry,
respectively.

For the three mixing angles, we allow the entire range, $0 \le \alpha_{1,2,3} \le 2 \pi$.
For $\tan\beta$ and $v_\Phi$, we use the following ranges
\begin{equation}
1\le \tan\beta \le  30 \quad
\textrm{and} \quad 
5\times 10^3 \le v_\Phi/\textrm{GeV} \le 10^4 .
\end{equation}
The ranges of the masses of the extra scalar bosons are taken to be $500 \le m_{H,A,H^\pm} /$GeV $\le 1500$,  
respectively, considering the bounds from flavor physics.
We have checked broader ranges for $\tan \beta$, $v_{\Phi}$, and the scalar boson masses, but 
we have found that it does not change the overall behavior of signal strengths in allowed regions.
We have also limited the angles in the ranges of
$ 0.5 \le \alpha_1 \le 1.5$, $1 \le \alpha_2 \le 2$,
and $1 \le \alpha_3 \le 2$ 
 because no allowed regions exist outside these ranges in the first scan with the entire range, 
or the other allowed regions does not change the overall feature.
At first, the Yukawa couplings of the extra quark and lepton are in the ranges of $ 0 \le y_{D,L} \le 4 \pi$,
respectively, 
but for simplicity we set $y_D = y_L = 1$.
We note that essential features from the analysis are not modified even though larger Yukawa couplings are chosen.  
The $U(1)_H$ gauge coupling must not be large because of constraints from the $Z'$ boson search 
at the LHC. We take $g_X$ to be in the range of $0.01 \le g_X \le 0.1$.
Then, the $Z'$ mass, $m_{Z'}$,  is written in terms of the parameters
\begin{equation}
m_{Z'} = \frac{g_H v}{2},
\end{equation}
where
\begin{equation}
g_H = \sqrt{ 4 g_X^2 \left( \sin^2\beta + \frac{v_\Phi^2}{v^2}\right)}.
\end{equation}
We find that the $Z'$ mass is in the range of $100 \sim 1000$ GeV for allowed parameters while the VEV for the $U(1)_H$ symmetry
breaking is in the range of $5 \sim 10$ TeV.

\begin{figure}[t]
\centering
\includegraphics[width=7cm]{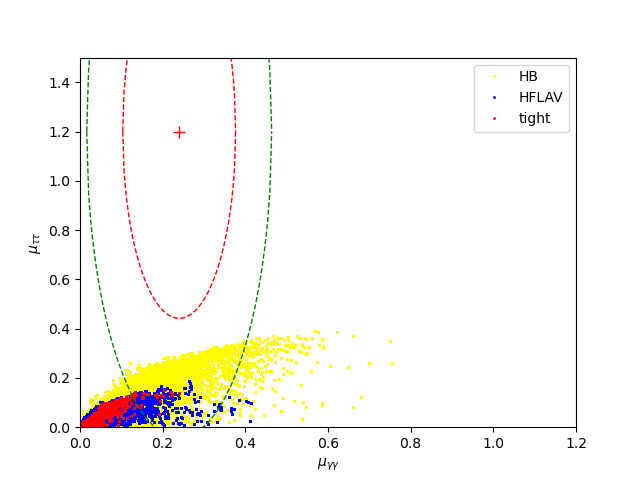}
\includegraphics[width=7cm]{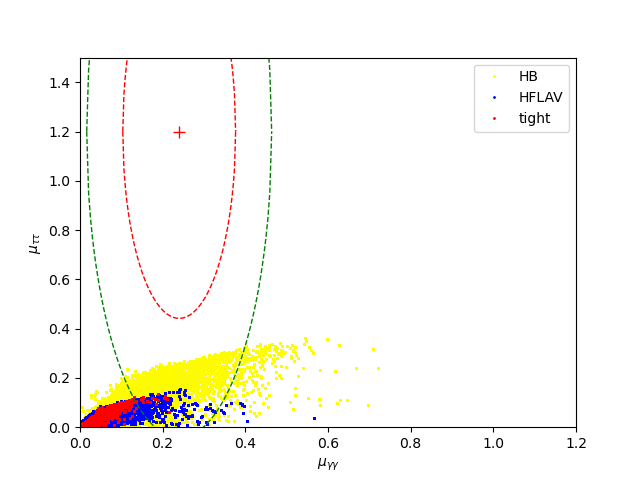}
\caption{
Signal strengths $\mu_{\gamma\gamma}$ and  $\mu_{\tau\tau}$ 
in the Type-II models with extra fermion contributions in the loop (left) and without them (right).
}\label{fig:type2-gg-tt}
\end{figure}

\begin{figure}[t]
\centering
\includegraphics[width=7cm]{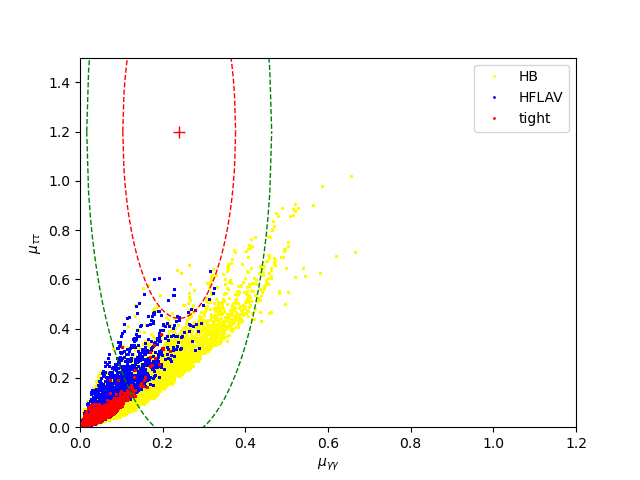}
\includegraphics[width=7cm]{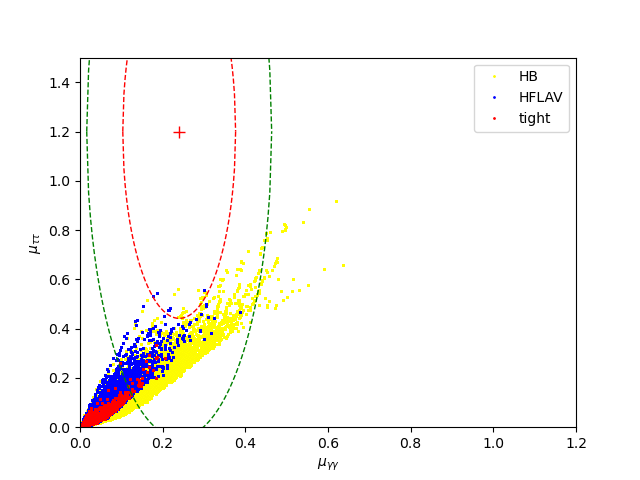}
\caption{
Signal strengths $\mu_{\gamma\gamma}$ and  $\mu_{\tau\tau}$ 
in the Type-Y models with extra fermion contributions in the loop (left) and without them (right).
}\label{fig:typeY-gg-tt}
\end{figure}

\subsection{Analysis}

We perform a scan of 11 parameters in the ranges described in the previous subsection.
Then, we apply the theoretical and experimental constraints to the parameter sets and 
calculate $\mu_{\gamma\gamma}, \mu_{\tau\tau}$ and $\mu_{bb}$ for those parameters
that pass those constraints.  

In Fig.~\ref{fig:type2-gg-tt}, we present  the signal strengths $\mu_{\gamma\gamma}$ and $\mu_{\tau\tau}$
in both Type-II models with extra fermion contributions in the loop (left) and without those contributions (right), respectively.
The extra fermion contribution in the loop can especially modify branching ratios of $\tilde{h} \to gg$ or $\gamma\gamma$. 
Thus, the production or decay rates of $\tilde{h}$ can be affected by the extra fermion loop contributions.
The two dashed lines denote $1\sigma$ and $2\sigma$ regions for the combined experimental values in ATLAS and CMS, respectively.
The blue and red points represent parameter sets consistent with the constraints discussed in the previous section. The only distinction between these two sets lies in the flavor physics constraints applied. 
The blue points satisfy the HFLAV cut for the $B\to X_s \gamma$ decays, which imposes a lower bound
of $m_{H^\pm}\gtrsim 500$ GeV.
Since the $B_s \to \mu^+ \mu^-$ constraint is more stringent than those from other observables
in both low and high $\tan\beta$ regimes~\cite{Li:2024kpd},
we impose the bound provided in Ref.~\cite{Li:2024kpd}.
The red points satisfy the tight cut, which is based on the constraints in Ref.~\cite{Li:2024kpd}.
We note that the lower limit of $m_{H^\pm}$ in this cut is about 800 GeV
while the $B_s \to \mu^+ \mu^-$ constraint is the same as those in the HFLAV cut.
Since the tight cut is more stringent than the HFLAV cut, all the red points overlap the blue points.
The yellow points represent parameter sets that satisfy all constraints except for those from HiggsSignals, while adhering to the HFLAV cut for the flavor physics constraints.
We note that the constraints from $t\bar{t}\tau^+ \tau^-$ production at the LHC are applied to both red and blue
points, while they are not to the yellow points.

We find that the expectation in our Type-II models is inconsistent with the experimental results within $1\sigma$, while
some regions are consistent within $2\sigma$. 
In particular, it is difficult to enhance the $\mu_{\tau\tau}$ 
because of effective Yukawa couplings of the 96 GeV scalar boson to 
the top quarks and tau lepton in the Type-II models:
\begin{equation}
    g_{\tilde{h}}^{t, \textrm{II}} = 
    \frac{\sin \alpha_1 \cos \alpha_2}{\sin \beta}, ~~~
   g_{\tilde{h}}^{\tau, \textrm{II}} =
    \frac{\cos \alpha_1 \cos \alpha_2}{\cos \beta},
    \label{effectiveYukawa2}
\end{equation}
which are normalized to the SM-like Yukawa couplings, respectively.
For the $gg\to \tilde{h} \to \tau\tau$ production, both Yukawa couplings are relevant.
Then, $g_{\tilde{h}}^{t, \textrm{II}}$ is proportional to
$\sin \alpha_1$, but $g_{\tilde{h}}^{\tau, \textrm{II}}$ is to $\cos\alpha_1$.
Therefore, it would be difficult to enhance both Yukawa couplings for the same $\alpha_1$.
It turns out that the branching ratio for $\tilde{h} \to \tau\tau$ is at most $0.1$ in the allowed parameter space,
which results in the small signal strength $\mu_{\tau\tau}$.

As we see in Fig. \ref{fig:type2-gg-tt}, 
the region allowed by the HFLAV cut is broader than that consistent with the tight cut, in the Type-II models. 
However, the total signal strengths do not differ significantly in the two cases.
We find that the constraint from HiggsSignals is very stringent, excluding  substantial portion of the parameter space. 
Without these constraints, the model prediction could deviate by up to approximately $1.3 \sigma$ from the experimental
central values.
This situation is similar in both models, with and without the extra fermion contributions in the loop. 
Despite this, it is apparent that the signal strengths are distinctive in both models.
This implies that a parameter set consistent within $2 \sigma$ with the extra fermion contributions might be excluded 
if the extra fermions are ignored in the loop, and vice versa.
Thus, the extra fermions may play a role in the search for the 96 GeV boson.

In Fig.~\ref{fig:typeY-gg-tt}, we present  the signal strengths $\mu_{\gamma\gamma}$ and $\mu_{\tau\tau}$
in both Type-Y models with extra fermion contributions in the loop (left) and without those contributions (right), respectively.
In these models, the effective Yukawa couplings are given by
\begin{equation}
\label{effectiveYukawaY}
    g_{\tilde{h}}^{t, \textrm{Y}} = g_{\tilde{h}}^{\tau, \textrm{Y}} = g_{\tilde{h}}^{t, \textrm{II}}.
\end{equation}
Since the two effective Yukawa couplings are identical,
both production cross sections for $gg\to \tilde{h} \to \gamma\gamma$
and $gg\to \tilde{h} \to \tau\tau$
tend to increase simultaneously as $\sin \alpha_1$ increases.
This behavior is indeed observed for both the blue and red points, which adhere to the HFLAV 
and tight cuts, respectively.
Moreover, compared to the Type-II models, the $\mu_{\tau\tau}$ signal strength can be enhanced 
and the deviation from the experimental central values can fall within the $1\sigma$ range under the HFLAV cut.
Even under the tight cut, the deviation can reach up to approximately $1.2\sigma$ from the central values.
As shown in Fig.~\ref{fig:typeY-gg-tt}, the effects of extra fermions in the loop are very restrictive,
but one can observe a small difference between two models with and without extra fermions in the loop.
We find that some points which are excluded the model without extra fermions could be consistent within $2\sigma$
by including the extra fermions in the loop, similar to the Type-II models.

\begin{figure}[t]
\centering
\includegraphics[width=7cm]{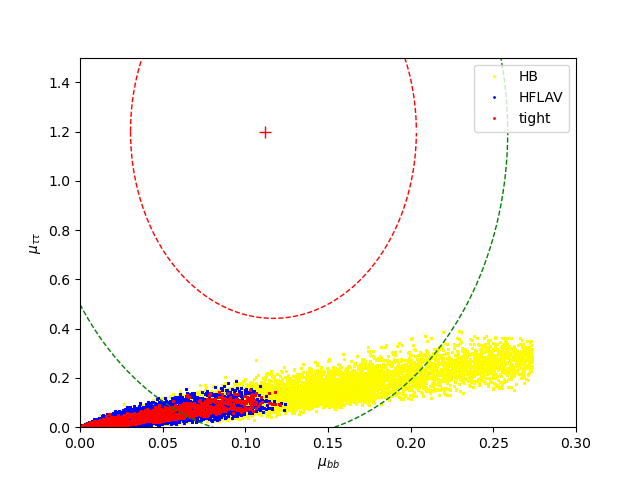}
\includegraphics[width=7cm]{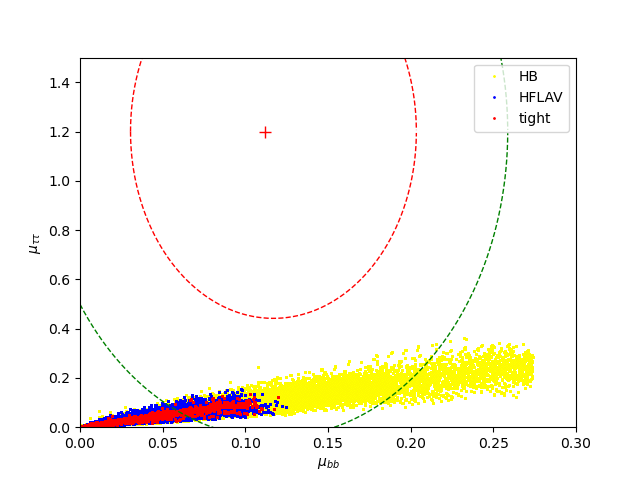}
\caption{
Signal strengths $\mu_{bb}$ and  $\mu_{\tau\tau}$ 
in the Type-II model with extra fermions in the loop (left) and without them (right).
}\label{fig:type2-bb-tt}
\end{figure}

\begin{figure}[t]
\centering
\includegraphics[width=7cm]{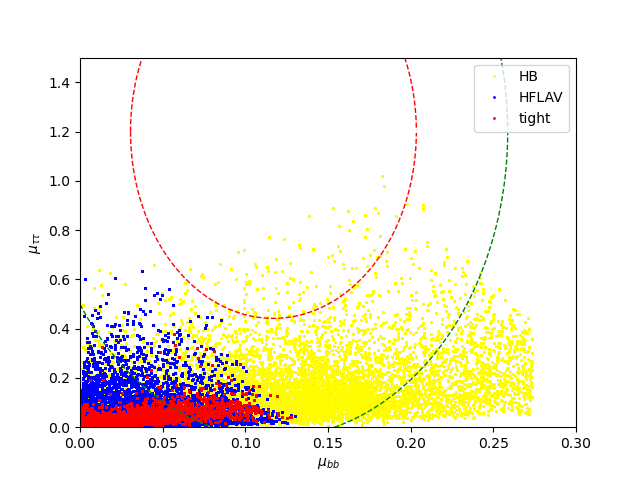}
\includegraphics[width=7cm]{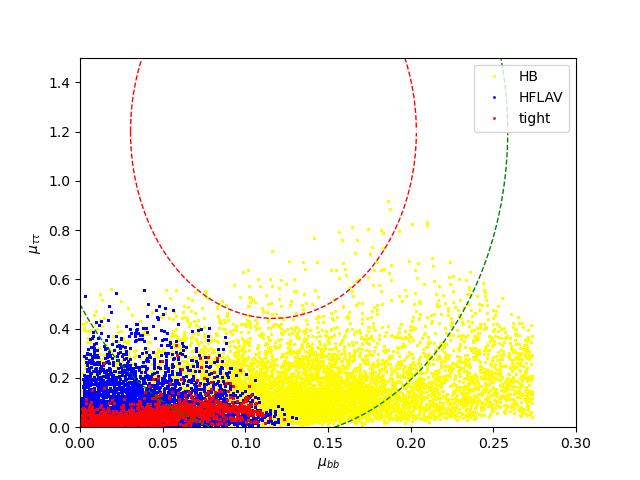}
\caption{
Signal strengths $\mu_{bb}$ and  $\mu_{\tau\tau}$ 
in the Type-Y model with extra fermions in the loop (left) and without them (right).
}\label{fig:typeY-bb-tt}
\end{figure}

In Figs.~\ref{fig:type2-bb-tt} and \ref{fig:typeY-bb-tt}, we present  the signal strengths $\mu_{bb}$ and $\mu_{\tau\tau}$
in the Type-II and Type-Y model 2HDMs, respectively.
The extra fermion contributions are included in the left figures, while they are not included in the right figures.
The effective couplings of $\tilde{h}$ to the bottom quark are 
\begin{equation}
 g_{\tilde{h}}^{b, \textrm{II}} = g_{\tilde{h}}^{b, \textrm{Y}} = g_{\tilde{h}}^{\tau, \textrm{II}}.
\end{equation}
In the Type-II 2HDM, the effective Yukawa coupling, $g_{\tilde{h}}^{b, \textrm{II}}$, of $\tilde{h}$ 
to the bottom quark is the same as
that to the tau lepton. Thus, $\mu_{bb}$ tends to be proportional to $\mu_{\tau\tau}$ as shown in the left figure of
Fig.~\ref{fig:type2-bb-tt}. It is obvious that $\mu_{bb}$ in the Type-II models is preferred to be similarly small like $\mu_{\tau\tau}$
in the allowed parameter space.
On the other hand, in the Type-Y models, the effective Yukawa coupling, $g_{\tilde{h}}^{b, \textrm{Y}}$, of $\tilde{h}$ 
to the bottom quark is different from that to the tau quark.
As shown in  Fig.~\ref{fig:typeY-bb-tt}, a wider region is allowed in the Type-Y models.

In our models, the extra fermions are required to ensure anomaly cancellation.
These fermions can interact with both gauge bosons and scalar bosons. 
As a result, the branching ratios of scalar bosons
into gauge bosons can be affected by extra fermions involving loop diagrams.
However, as shown in the figures in this section, the overall features of the models remain unchanged
by the inclusion of extra fermion contributions.
Therefore, the impact of the extra fermions appears to be insignificant for the 96 GeV resonances.
This is because the parameters in the scalar potential are tightly constrained by experimental data,
such as flavor physics observables and precision measurements of the SM Higgs boson.
The extra fermions play only a minor role in both types of experiments.
However, we note that certain detailed features of the models may differ depending on 
whether the extra fermions are included. 
The signal strengths, $\mu_{\gamma\gamma}$ and $\mu_{\tau\tau}$
depend on the branching ratios of  $\tilde{h}$ decays into $gg$ and/or $\gamma\gamma$.
Therefore, the extra fermion contributions to these branching ratios may become significant 
if a given parameter point lies near the boundary of the $2\sigma$ allowed region.

\subsection{Benchmark points}
\label{sec;benchmark}

In Table \ref{benchmark}, we introduce some benchmark points that explicitly show our predictions. 
Those parameter sets evade all constraints discussed in Sect. \ref{sec3}. 
We can see  both cases: the heavy charged Higgs case and the relatively light charged Higgs case in the Type-II and the Type-Y 2HDMs, respectively.  
In all cases, $\mu_{\gamma \gamma}$, $\mu_{\tau \tau}$ and $\mu_{bb}$ are within or close to the 2$\sigma$ ellipse.
$\mu_{\tau \tau}$ in the Type-Y tends to be larger than that in the Type-II, because of the mixing angle dependence.

There is no big difference between in the Type-II 2HDM and in the Type-Y 2HDM, when charged Higgs is heavy, i.e., $m_{H^\pm}>800$ GeV.
We note that the parameter values of point 1 in Table \ref{benchmark} are the same as them of point 3. We can see the difference of the mixing dependence on the $\tau$ couplings
in the signal strengths. The mixing angles, $\alpha_1$, $\alpha_2$ and $\alpha_3$, in all cases,
correspond to ${\cal O} (1)$ $|\sin \alpha_{1,2,3}|$. Then,
$\tilde{h}$, whose mass is 96 GeV, is dominated by $h_\Phi$,
while the SM-like Higgs boson, $h$, dominantly consists of $h_1$ and $h_2$.
In all points in Table \ref{benchmark}, $\tan \beta$ is small, and $|\cos \alpha_{1,2}|$ are larger than $0.1$ to enhance $\mu_{\gamma \gamma}$ and $\mu_{\tau \tau}$.
As shown in Table \ref{benchmark},
$g_X$ is very small to evade too large deviation of the $\rho$ parameter,
so, the mixing parameters are relevant to the signal strengths.

\begin{table}[h]\centering
	\begin{tabular}{|c||c|c || c|c|}\hline

   & point 1 & point 2 & point 3 & point 4 \\  \hline 
Model & Type-II  &  Type-II &  Type-Y &  Type-Y \\      \hline \hline 
$\alpha_1$  &  1.01   &  0.97   &  1.01   & 1.36    \\ \hline    
     $\alpha_2$  & 1.88  & 1.25 &  1.88  & 1.90  \\ \hline    
       $\alpha_3$  & 1.40    & 1.48   &  1.40  & 1.30       \\ \hline
         $\tan \beta$  & 1.20 & 1.21  & 1.20 & 1.98    \\ \hline 
       $v_\Phi$ (GeV) & 7459.47  & 8646.47 &   7459.47  &  9899.79    \\ \hline   
       $m_H$ (GeV)  &    900.36  & 749.49  & 900.36 &  652.26   \\ \hline
       $m_A$ (GeV)  & 713.55 &  630.49 &  713.55  & 549.93   \\ \hline  
      $m_{H^\pm}$ (GeV) &  877.49   & 692.21 &   877.49 & 605.38  \\ \hline        
         $y_D$  &  1.0    &  1.0   &  1.0   &   1.0   \\ \hline           
         $y_L$  &   1.0   &   1.0  &   1.0  &   1.0    \\ \hline 
          $g_X$  &  $   3.83  \times 10^{-2}$ & $ 7.29  \times 10^{-2}$ & $ 3.83   \times 10^{-2}$ & $  1.61  \times 10^{-2}$ \\ \hline  
          \hline \hline        
         $\mu_{\gamma \gamma}$   & 0.13 &  0.15  & 0.13 & 0.27   \\ \hline
            $\mu_{\tau \tau}$  &  $9.07 \times 10^{-2}$   & 0.13  & 0.15 & 0.35   \\ \hline      
    $\mu_{bb}$  &   $ 8.59  \times 10^{-2}$ & $ 9.25  \times 10^{-2}$  &    $ 8.08 \times 10^{-2}$    & $ 5.23 \times 10^{-2}$  \\ \hline   
	\end{tabular}
	\caption{\label{benchmark} Benchmark points in the Type-II and Type-Y 2HDMs with $U(1)_H$ gauge symmetry.
	}
\end{table}

\section{Summary}
\label{sec5}

We studied the di-photon and di-$\tau$ channels at the LHC as well as the $b \bar{b}$ channel at  LEP, 
in the Type-II and Type-Y 2HDMs with gauged $U(1)_H$ symmetry. 
We can simply add one extra Higgs doublet to the SM. Such a simple extension, however, causes
FCNCs at the tree level, if two Higgs doublets are not distinguished. Moreover, extra scalars predicted by
the two Higgs doublets usually reside around the EW scale, so that the extended SMs are easily excluded.
In the conventional 2HDMs (e.g., the Type-II 2HDM and the Type-Y 2HDM), softly broken $Z_2$ symmetry is
imposed on fields and the strong constraints are evaded. In our model, we see that softly broken $Z_2$ symmetry is originated from the spontaneous symmetry breaking of gauged $U(1)_H$ symmetry. In this sense, 
our model can be interpreted as the underlying theory of the conventional 2HDMs.
Moreover, the 2HDM with a singlet scalar is also recently studied, motivated by the excesses around 96 GeV.
Our 2HDM with gauged $U(1)_H$ symmetry also predicts a singlet scalar, so it is worthwhile to study the signals related to the resonance around 96 GeV in our models. In our model,
the $U(1)_H$ gauge symmetry distinguishes 
one Higgs doublet from the other, and naturally suppresses tree-level FCNCs.
One extra scalar charged under $U(1)_H$ is also introduced, so that there are three CP-even 
neutral scalars at low energy. In our study, the mass of lightest neutral scalar, denoted by $\tilde{h}$, was fixed at 96 GeV that corresponds to the mass region where the mild excesses are reported by the ATLAS and CMS collaborations. In our models, extra fermions as well as extra scalars
contribute to the di-photon and di-$\tau$ channels. The extra gauge boson deviates $\rho$ parameter
from $1$. We surveyed our predictions of the signal strengths, $\mu_{bb}$, $\mu_{\gamma \gamma}$ and 
$\mu_{\tau \tau}$, scanning parameters. The  $Z^\prime$ couplings are strongly constrained by
the $\rho$ parameter and the $Z^\prime$ searches, while the contribution of extra fermions affects the signal strengths
in some parameter sets. We saw that the dominant contributions to $\mu_{\gamma \gamma}$ and $\mu_{\tau \tau}$ are from the mixing among three neutral scalars, not from the extra new fermions.
As discussed in Sect. \ref{sec4}, the lightest neutral scalar is mainly composed of the scalar from $\Phi$. This result is consistent with the previous work \cite{Biekotter:2023oen}.
Our work reveals the underlying theory of the 2HDM with a complex scalar field \cite{Biekotter:2023oen}, and explicitly shows how large contributions of the extra fields 
predicted by extending the symmetry, that distinguishes two Higgs fields, to gauged $U(1)_H$ symmetry can be. The predictions of the signal strengths, in fact, deviate from them of
the 2HDMs without extra fermions. Further study would be required to reveal
which setup is preferred by the excesses around 96 GeV.
The improved SM prediction of $B \to X_s \gamma $ may exclude the parameter region with $m_{H^\pm} < 800$ GeV \cite{Misiak:2020vlo}, 
but the new HFLAV result \cite{HeavyFlavorAveragingGroupHFLAV:2024ctg}
would relax the bound drastically. We plot our predictions in both heavy and light charged Higgs cases.
There is no big difference between in the Type-II 2HDM and in the Type-Y 2HDM, when charged Higgs is heavy. If the charged Higgs is light,
the signal strength, $\mu_{\tau \tau}$, tends to be larger in the Type-Y than that in the Type-II. 
In our model, there are extra quarks which can contribute to $B \to X_s \gamma $. The study of flavor physics involving extra quarks, however, is not simple, since  other flavor observables are usually correlated to $B \to X_s \gamma $. This is also our future work. 
Finally, let us comment on how to distinguish our models with the others in the previous works. 
As we see in Table \ref{benchmark}, $Z^\prime$ mass, that is ${\cal O}(g_X v_\Phi) $, is below 1 TeV in our model. There are several possibilities of $U(1)_H$ charge assignments and matter fields as discussed in Ref. \cite{Ko:2012hd}, but in any case, the resonance search below 1 TeV play an important role in testing our models. As studied in Ref \cite{Ko:2015fxa}, there is a good DM candidate in the Type-II 2HDM with gauged $U(1)_H$ symmetry. Our analysis in this paper does not include the DM, but 
this prediction is also important to test and distinguish our model with the others.

\acknowledgments

This work is supported in part by National Research Foundation of Korea (NRF) Grant No. NRF-2018R1A2A3075605 and No. RS-2023-00270569 (S.\,B.), KIAS Individual Grants under Grant No. PG021403 (P.\,K.),
Grant-in-Aid for Scientific research from the MEXT, Japan, No.\,24K07031 (Y.\,O.), 
Basic Science Research Program through the National Research Foundation of Korea (NRF)
funded by the Ministry of Science, ICT, and Future Planning under the Grant No. NRF-2021R1A2C2011003m 
and RS-2023-00237615 (C.\,Y.).


\appendix


\section{Bounded-from-below condition}
\label{bounded}

We present constraints from the condition for scalar potential to be bounded from below.
For the derivation of the constraints, we follow the method derived in Ref. \cite{Horejsi:2005da,Muhlleitner:2016mzt}.
The quartic couplings are constrained by
\begin{eqnarray}
\lambda_{1,2,\Phi}>0,\quad  \lambda_3+(\lambda_4\pm \lambda_4)/2 + 2\sqrt{\lambda_1\lambda_2}>0,\quad
\tilde{\lambda}_{1,2}+2\sqrt{\lambda_{1,2}\lambda_\Phi}>0
\nonumber \\
\end{eqnarray}
together with the following four conditions as follows
\begin{eqnarray}
&&[(\lambda_3>0 \cap \lambda_{34}>0) \cup (E_1)]
\cap [(\tilde{\lambda}_1>0 ) \cup (4\lambda_1\lambda_\Phi - \tilde{\lambda}_1^2>0)]
\nonumber \\
&\cap &[(\tilde{\lambda}_2>0 ) \cup (4\lambda_2\lambda_\Phi - \tilde{\lambda}_2^2>0)]
\cap [(\tilde{\lambda}_1>0 \cap \tilde{\lambda}_2>) \cup (E_2)],
\end{eqnarray}
where
\begin{eqnarray}
E_1 &=&
[\lambda_{12}^2 - \lambda_3^2 >0]
\cap [\lambda_{12}^2 - \lambda_{34}^2 >0]
\cap [\lambda_{12}^2 - \lambda_3 \lambda_{34}
+\sqrt{(\lambda_{12}^2-\lambda_3^2)(\lambda^2-\lambda_{34}^2)}>0],
\\
E_2 &=&
[\lambda_{1\Phi}^2 - \tilde{\lambda}_1^2 >0]
\cap [\lambda_{2\Phi}^2 - \tilde{\lambda}_2^2 >0]
\cap [2(\lambda_3+D)\lambda_\Phi- \tilde{\lambda}_1\tilde{\lambda}_2
+\sqrt{(\lambda_{1\Phi}^2-\tilde{\lambda}_1^2)(\lambda_{2\Phi}^2-\tilde{\lambda}_2^2)}>0].
\end{eqnarray}
Here,
\begin{eqnarray}
\lambda_{12}^2&=&4\lambda_1 \lambda_2, \quad
\lambda_{1\Phi}^2=4\lambda_1 \lambda_\Phi, \quad
\lambda_{2\Phi}^2=4\lambda_2 \lambda_\Phi,
\\
D&=& \min (\lambda_4,0).
\end{eqnarray}

\section{Effective couplings}
\label{effcoup}

In this appendix, we present the definition of the effective couplings for the triple vertices of gauge bosons and scalar bosons.
The effective couplings are defined by the couplings for the triple vertices normalized by those in the SM
\begin{eqnarray}
g_{W^{\pm}H^{\mp} A} &=&
(V_A)_{33}\cos\beta - (V_A)_{32} \sin\beta,
\\
g_{W^{\pm}H^{\mp} S_i} &=&
(R)_{i2}\cos\beta - (R)_{i1} \sin\beta,
\\
g_{Z A S_i} &=&
\cos\theta \left[ (R)_{i1} (V_A)_{32} + (R)_{i2} (V_A)_{33} \right]
\nonumber \\
&-&
\frac{2 g_X}{g_2} \sin\theta
\left[
(R)_{i2} (V_A)_{33} - (R)_{i3} (V_A)_{31}
\right],
\\
g_{Z Z S_i} &=&
\cos^2 \theta 
\left[ (R)_{i1} \cos\beta + (R)_{i2} \sin\beta \right]
-
\frac{4 c_W^2 g_X^2}{g_2^2} \sin\theta (R)_{i2}  \sin\beta
\nonumber \\
&+&
\frac{4 c_W^2 g_X^2}{g_2^2}  \sin^2\theta  
\left[ (R)_{i2} \sin\beta + (R)_{i3} \frac{v_\Phi}{v} \right],
\\
g_{Z Z' S_i} &=&
\frac{2 c_W g_X}{g_2}
(R)_{i2} \sin\beta ( \cos^2 \theta -  \sin^2\theta)
\nonumber \\
&+&
\frac{\sin 2\theta}{2}
\left[ (R)_{i1} \cos\beta + (R)_{i2} \sin\beta 
+\frac{4 c_W^2 g_X^2}{g_2^2}
\left\{
(R)_{i2} \sin^2\beta + \frac{v_\Phi}{v} (R)_{i3}
\right\}
\right],
\\
g_{W W S_i} &=&
{(R)_{i1} \cos\beta +(R)_{i2} \sin\beta},
\\
g_{W^\pm H^\mp Z} &=&
\frac{g_X}{g_2} \sin\theta \sin 2\beta,
\\
g_{W^\pm H^\mp Z'} &=&
-\frac{g_X}{g_2} \cos\theta \sin 2\beta,
\end{eqnarray}
where $S_i \equiv (\tilde{h}, h, H)$.


\end{document}